\documentclass[letters,usenatbib]{mnras}
\usepackage{newtxtext,newtxmath,graphicx}

\usepackage[T1]{fontenc}

\DeclareRobustCommand{\VAN}[3]{#2}
\let\VANthebibliography\thebibliography
\def\thebibliography{\DeclareRobustCommand{\VAN}[3]{##3}\VANthebibliography}

\newcommand\src{AT 2018fyk}
\newcommand\xmm{\textit{XMM-Newton}}
\newcommand\obsid{0853980201}
\newcommand\frezero{6.51\times10^{-5}~\rm Hz}

\date{Accepted XXX. Received YYY; in original form ZZZ}

\title[Soft lags in AT2018fyk]{XMM-Newton detection of soft time lags in the TDE candidate AT 2018fyk}

\author[Wenda Zhang]{Wenda Zhang\thanks{E-mail: wdzhang@nao.cas.cn}\\
National Astronomical Observatories, Chinese Academy of Sciences,
20A Datun Road, Beijing 100101, China}

\begin{document}
\label{firstpage}
\pagerange{\pageref{firstpage}--\pageref{lastpage}}
\maketitle

\begin{abstract}
In this letter we report a tentative detection of soft time lags (i.e. variability of softer photons lags behind the variability of harder photons)
in one \xmm{} observation of the tidal disruption event (TDE) candidate \src{}
while the source was in the hard spectral state. The lags are detected
at $\frezero$. The amplitude of the lags with respect to 0.5$-$1 keV monotonically decreases with the photon energy,
from $\sim 1200~\rm s$ at 0.3$-$0.5 keV to $\sim -4200~\rm s$ at 3$-$5 keV (in our convention a positive lag means lagging behind the
reference band). We find that the amplitude is proportional
to the logarithm of the energy separation between the examined band and the reference band.
The energy-dependent covariance spectrum indicates that the correlated variability is more likely to be associated with the non-thermal radiation.
The soft lags are difficult to reconcile with the reverberation 
scenario that are used to explain the soft lags in active galactic nuclei.
On the other hand, the observed soft lags are consistent with the picture that
the soft X-rays are down-scattered hard X-rays by the outflow as predicted by ``unification'' models of TDEs.

\end{abstract}
\begin{keywords}
transients: tidal disruption events -- accretion, accretion discs -- black hole physics
\end{keywords}

\section{Introduction}

Tidal disruption events (TDEs) are transient events caused by super-massive black holes (SMBHs) tidally disrupting nearby stars and 
accreting stellar debris \citep[see][for a review]{gezari_tidal_2021}. TDEs offer us unique opportunities to study
dormant SMBHs. So far, a few tens of TDEs have been discovered.
Depending on their X-ray properties, TDEs can be classified into two categories: X-ray TDEs that are detected in X-rays,
and optical-UV TDEs that lack X-ray emission. Understanding the X-ray/optical-UV dichotomy is an important
task in the study of TDEs.

To explain the weak X-ray emission of optical-UV TDEs, several models have been proposed.
The first set of models (often referred to as ``unification'' models of TDEs) claimed that the 
dichotomy is a viewing angle effect:
the X-ray and optical-UV TDEs are intrinsically the same, but for observers
at large inclinations the X-ray radiation is obscured by the wind launched by the super-Eddington accretion disc
\citep[e.g.][]{dai_unified_2018}. In the second set of models, the dichotomy is intrinsic; the strength of the X-ray emission
depends on the physical properties of the system. 
For instance, it has been proposed that the disc formation is faster for
TDEs with larger black hole mass \citep[e.g.][]{guillochon_dark_2015},
resulting in the absence of X-ray emission in the early outburst stage of low-mass TDEs.

\src{} is a candidate X-ray TDE discovered by the ASAS-SN on September 8th, 2018 \citep{stanek_asas-sn_2018}.
It is located at the nucleus of LCRS B224721.6$-$450748, a galaxy at a redshift of 0.06\footnote{https://ned.ipac.caltech.edu/byname?objname=LCRS+B224721.6-450748}.
\citet{wevers_fainter_2020} measured the mass of the central SMBH 
of the galaxy using the $M-\sigma$ relation, and found
${\rm log_{10}}\ M_{\rm BH}=7.7\pm0.4$ where $M_{\rm BH}$ is the black hole mass in solar mass.
\citet{wevers_evidence_2019}
performed a multi-wavelength study of \src{} and found evidences for rapid disc formation. 
\citet{wevers_rapid_2021} studied the long-term evolution of the X-ray emission of \src{},
and discovered an X-ray state transition of \src{} from a soft, thermal-dominated spectral
state to a hard, non-thermal dominated state. The spectral state transition is accompanied by
a change in the timing properties, with 
high-frequency variability ($\gtrsim 10^{-5}~\rm Hz$) emerging after the state transition.

We perform Fourier time-lag analysis of \src. 
Fourier time-lag analysis has been proved to be a great tool to study the accretion flow around black holes with mass
across a large dynamic range. Time-lag analysis of both active galactic nuclei (AGNs) 
and black hole X-ray binaries (BHXRBs) has revealed continuum hard lags \citep[e.g.][]
{papadakis_frequency-dependent_2001,mchardy_combined_2004,arevalo_spectral-timing_2006,arevalo_X-ray_2008,
sriram_energy-dependent_2009,epitropakis_X-ray_2017,papadakis_long_2019,miyamoto_delayed_1988,miyamoto_X-ray_1989,
nowak_phase_1996,nowak_rossi_1999,cui_phase_2000, crary_hard_1998,reig_energy_2003,altamirano_evolution_2015}.
On top of the continuum hard lags, there have been detections of soft lags where the variability of softer bands are lagging behind the
harder bands.
The soft lags were first detected by \citet{fabian_broad_2009} in 1H 0707$-$495, and then in other AGNs \citep[e.g.][]{zoghbi_X-ray_2011,
cackett_soft_2013,fabian_long_2013,de_marco_discovery_2013,alston_X-ray_2014}, as well as in BHXRBs
\citep[e.g.][]{uttley_causal_2011,de_marco_tracing_2015,de_marco_evolution_2017}. The soft lags are believed to be the lags between
the hard coronal emission and the reflection emission due to the hard coronal emission irradiating the disc, and are consequently referred
to as ``reverberation lags''. \citet{de_marco_discovery_2013} performed a systematic study
of the reverberation lags, and found correlations between the amplitude of the reverberation lags and
the black hole mass, and between the variability frequency and the black hole mass.

Soft lags have been detected in ultra-luminous X-ray sources (ULXs) as well \citep{heil_linear_2010,de_marco_time_2013,
hernandez-garcia_X-ray_2015,pinto_ultraluminous_2017,kara_discovery_2020,pintore_rare_2021,mondal_evidence_2021}.
ULXs are extragalactic, off-nuclear, ultra-luminous ($\gtrsim 10^{39}~\rm erg~s^{-1}$) X-ray sources.
It is generally believed that the majority of ULXs are stellar-mass black holes or neutron stars accreting at super-Eddington
rates, instead of intermediate-mass black holes (IMBHs) accreting at sub-Eddington rates.
Most of the soft lags in ULXs have amplitudes of a few hundreds to thousands of seconds and are usually interpreted as
the time lags due to hard X-ray photons getting down-scattered by the outflow.


In this letter we report a detection of soft lags in \src.
The letter is organised as follows. The procedure of data reduction is illustrated in Sec.~\ref{sec:data_reduction}.
The results are presented in Sec.~\ref{sec:results}.
We discuss the results in Sec.~\ref{sec:discussion}.

\section{Data Reduction}
\label{sec:data_reduction}

\xmm{} observed \src{} three times. The source was detected in the first (obs id: 0831790201) and second (obs id: \obsid{}) observations, and
undetected in the last observation.
During the first observation \src{}
was in the soft state, and no X-ray variability was detected above $\sim 10^{-5} ~\rm Hz$, while during the second observation strong high-frequency
($\gtrsim 10^{-5}~\rm Hz$) X-ray variability was discovered \citep{wevers_rapid_2021}.
We therefore perform Fourier time-lag analysis on the pn data of \obsid{}.

The observation \obsid{} were taken on Oct. 27th, 2019. During the observation pn was operating in the full-frame mode.
The data are reduced with the the \xmm{} science analysis system (SAS) version 18
following standard procedures\footnote{https://www.cosmos.esa.int/web/xmm-newton/sas-threads}.
We produce the pn clean event list with \textsc{epproc}. We filter the event
list to remove periods contaminated by high particle background flares, where the good time interval is defined to be the period during 
which the 10$-$12 keV count rate is no larger than $0.4~\rm counts/s$\footnote{https://www.cosmos.esa.int/web/xmm-newton/sas-thread-epic-filterbackground}.
We are left with a $\sim 30~\rm ks$ exposure after removing flaring backgrounds.
The source and background lightcurves are extracted from the filtered
event list with \textsc{evselect}. When extracting the lightcurves we select single and double events ($PATTERN\leq 4$) with $FLAG==0$.
The source region is a circle centering on the source position with a radius of 35''. The background lightcurves are extracted from a 
source-free region close to the source with the same radius. The source lightcurves are then corrected for background and various other effects
using \textsc{epiclccorr}.
We also extract source and background spectra with \textsc{evselect}, and generate corresponding ancillary file and redistribution matrix using
\textsc{arfgen} and \textsc{rmfgen}, respectively.

We extract 30-s lightcurves of 5 energy bands: $0.3-0.5~\rm keV$, $0.5-1~\rm keV$, $1-2~\rm keV$, $2-3~\rm keV$, and
$3-5~\rm keV$. Above $\sim 5$ keV the background is stronger than the source so we exclude the $>5~\rm keV$ bands in our analysis.
For each energy band we split the lightcurve into 2 segments, each $\sim 15~\rm ks$ long. Taking $0.5-1~\rm keV$ as the reference, we 
follow \citet{epitropakis_statistical_2016} to compute the frequency-dependent Fourier time lags.
We estimate the significance and uncertainty of the time-lag measurements by performing Monte$-$Carlo simulations.
Detailed procedures of the simulations can be found in the Appendix.


\section{Results}
\label{sec:results}
\subsection{Time-lag analysis}
\begin{figure}
\includegraphics[width=\columnwidth]{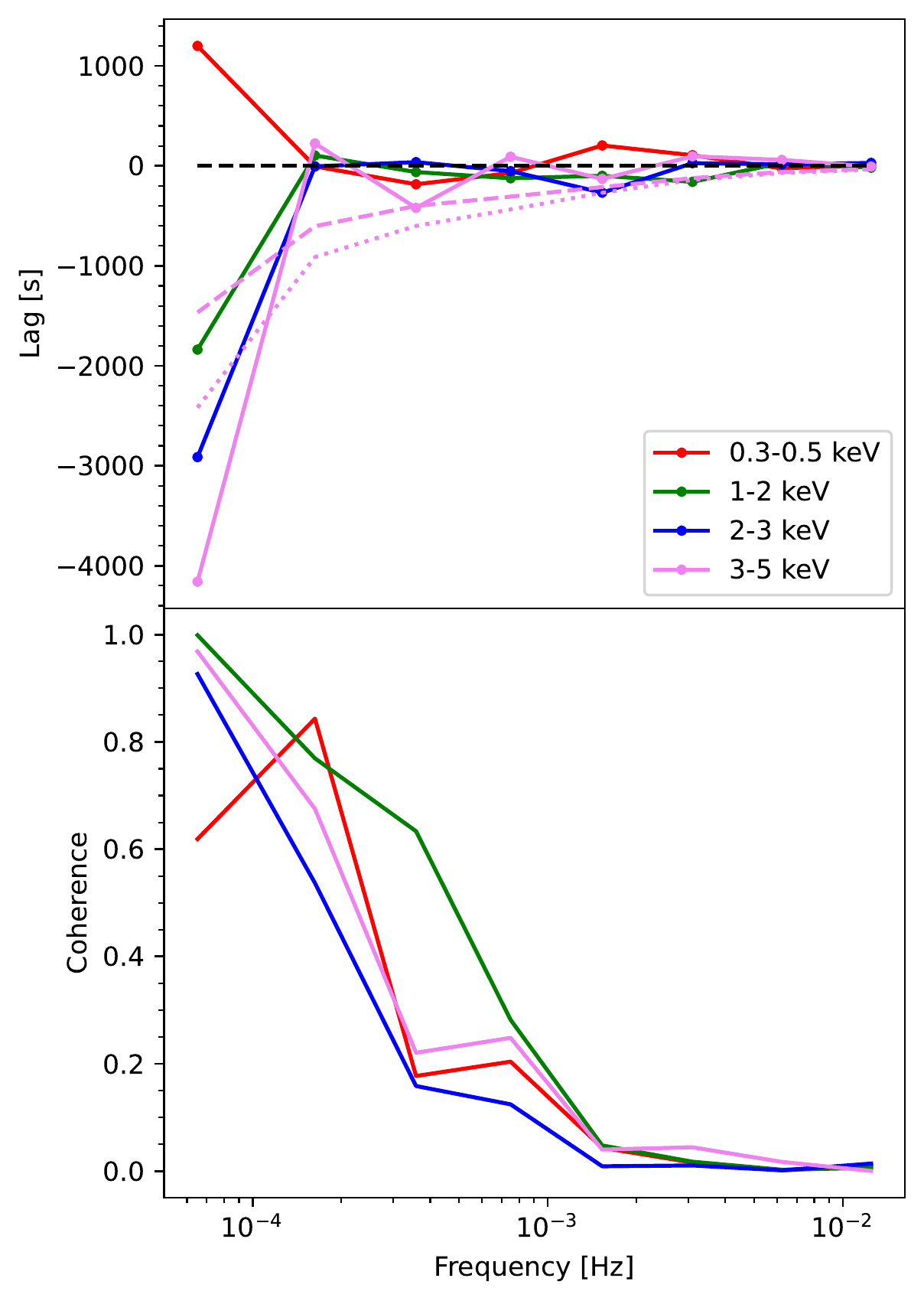}
\caption{The Fourier time lags (upper panel) with respect to $0.5-1~\rm keV$ and the coherence (lower panel).
The results of different energy bands are plotted in different colors, as indicated in the plot.
In our convention a positive lag means lagging behind the reference band.
For $3-5~\rm keV$, the 90\% and 95\% confidence levels are plotted in dashed and dotted lines, respectively.
\label{fig:lag}}
\end{figure}

The frequency-resolved time lags and the coherence are presented in the upper and lower panels of Fig.~\ref{fig:lag}, respectively.
Note that in our convention a positive lag means lagging behind the reference band.
The coherence is close to unity at the lowest frequency $\frezero$, 
decreases with the frequency, and drops to zero above $\sim2\times10^{-3}~\rm Hz$.

The time lags are close to zero except at $\frezero$, the lowest frequency we can reach.
At $\frezero$, the confidence levels of the time-lags are $92.0\%$, $97.5\%$,
$97.9\%$, and $97.9\%$, for 0.3$-$0.5 keV, 1$-$2 keV, 2$-$3 keV, and 3$-$5 keV, respectively.
The maximum confidence level of $97.9\%$ corresponds to a significance of $\sim 1.8\sigma$.
The most surprising result is that the time lags are soft,
i.e. variability of soft photons are lagging behind variability of hard photons.
This is opposite to AGNs where the continuum time lags are hard
\citep[e.g.][]{arevalo_spectral-timing_2006,arevalo_X-ray_2008,sriram_energy-dependent_2009,epitropakis_X-ray_2017}.
The amplitude of the time lags monotonically decreases with the energy, from $1200~\rm s$ at the 0.3$-$0.5 keV band to
$-4200~\rm s$ at the 3$-$5 keV band. 

\begin{figure}
 \includegraphics[width=\columnwidth]{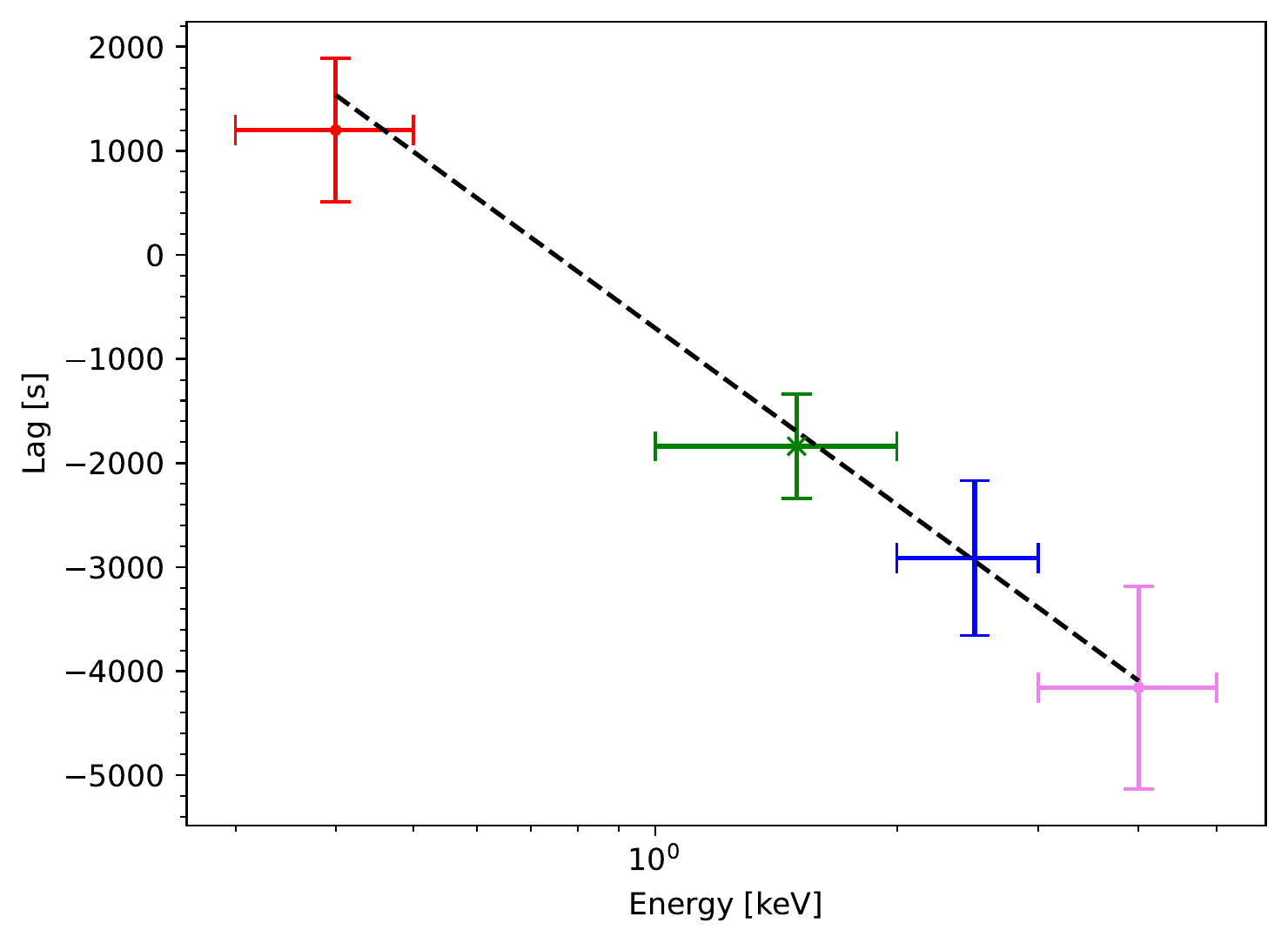}
 \caption{The time lags at $\frezero$ as a function of the central energy of the energy bands.
 The error bars correspond to 68\% uncertainty.
 We fit the lag-energy data with a logarithmic
 function and the best-fit model is shown by the black dashed line.\label{fig:lagvse}}
\end{figure}

In Fig.~\ref{fig:lagvse} we plot the amplitude of the time lags at $\frezero$ as a function of the centroid energy of the energy bands. 
The data points seem to fall on a straight line on the plot where the time lags and energy are plotted in linear and logarithmic scales,
respectively. This implies that the time lags depend logarithmically on the energy separation between two energy bands.
We fit the data with a logarithmic model, and find:
\begin{equation}
\label{eq:logfit}
\tau = (-5.63\pm0.80)\times10^3\ {\rm log_{10}}\left(\frac{E_2}{E_1}\right)~\rm s, 
\end{equation}
where $\tau$ is the amplitude of the time lag, and $E_1$, $E_2$ are the centroid energies of the two energy bands.
The best-fit model is plotted in a dashed line in Fig.~\ref{fig:lagvse}. 
By comparing the data and the model, it is obvious that the model fits the data well, confirming the logarithmic dependence.

\subsection{Energy and covariance spectrum}
\begin{figure}
 \includegraphics[width=\columnwidth]{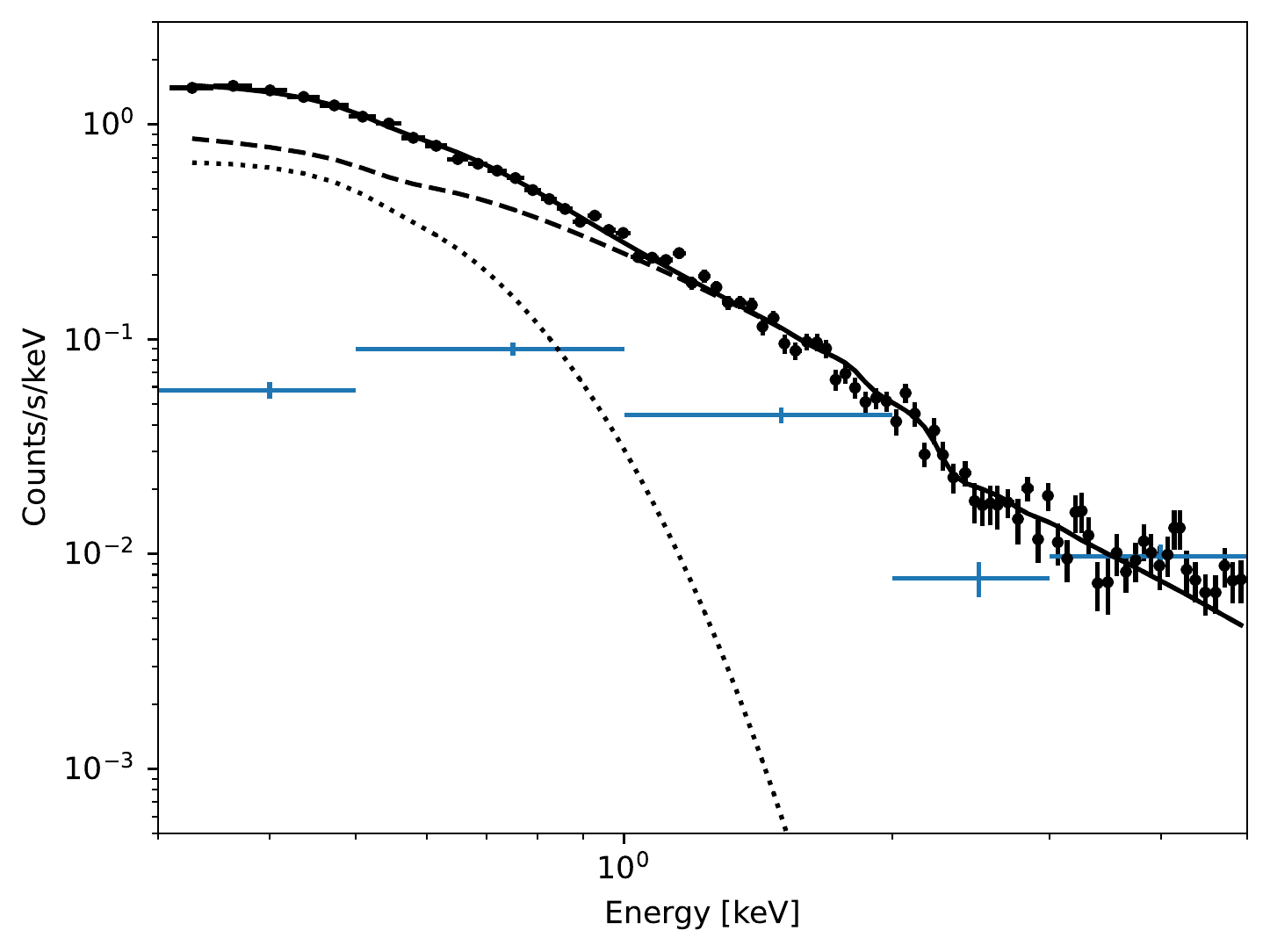}
 \caption{The covariance spectrum in blue colors. We also plot the energy spectra (black dots), the best-fit model (black solid), and
 the two components of the model: blackbody (black dotted) and powerlaw (black dashed).\label{fig:covspec}}
\end{figure}

We fit the pn energy spectrum with Xspec 12.11.1. The spectrum is rebinned with \textsc{specgroup} to ensure minimum
counts of 30 per bin. We fit the model to the data using the $\chi^2$ statistics. We fit the pn spectrum with the following spectral model:
$phabs*(blackbody+powerlaw)$, which includes two components: a thermal blackbody component, and 
a nonthermal powerlaw component, both of which are absorbed by interstellar media.
The spectrum and best-fit model are plotted in Fig.~\ref{fig:covspec}.
The best-fit model (black solid) yields $\chi^2_\nu=1.47$ (with 83 degrees of freedom), and consists of 
a dominating powerlaw component (black dashed) with $\Gamma= 2.32^{+0.09}_{-0.07}$ and
a soft thermal component (black dotted) with a temperature of $100^{+7}_{-4}~\rm eV$.

To investigate the origin of the correlated variability, we also calculate the covariance spectrum in the frequency range of
$3.7\times10^{-5}-1.7\times10^{-4}~\rm Hz$. 
We extract lightcurves with time resolution of $3000~\rm s$, and split the lightcurves into 2 segments, each of which contains 9 bins.
Then we compute the covariance spectrum and its uncertainty 
following \citep{wilkinson_accretion_2009}, taking the 0.5$-$1 keV band as the reference.
The covariance spectrum is plotted in blue in Fig.~\ref{fig:covspec}. The shape of the covariance spectrum indicates that the
correlated variability is more likely to be associated with the non-thermal emission than the thermal emission.

\section{Discussion and Conclusions}
\label{sec:discussion}

We tentatively detect soft time lags in one \xmm{} observation of the TDE candidate \src{}.
The observation (obs id: \obsid{}) was taken while the source was in the hard spectral state.
The lags are detected at $\frezero$, and the amplitude of the lags with respect to the 0.5$-$1 keV reference band are proportional
to the logarithm of the energy separation between the examined band and the reference band, ranging from 
$1200$ to $-4200$ seconds. The time-lags of 2$-$3 keV and 3$-$5 keV are most significant, both with a confidence level of
$97.9\%$.

A model fit confirms the apparent logarithmic dependence. The coherence is close to unity at $\frezero$ and drops rapidly
with the variability frequency.

We extract the energy spectrum and fit it with a two-component model. The best-fit model consists of a thermal component
with a temperature of $\sim 100~\rm eV$ and a dominating non-thermal component with a photon index of $\sim 2$.
We also calculate the energy-dependent covariance spectrum. The shape of the covariance spectrum indicates that the
correlated variability is associated with the non-thermal emission.

Below we discuss possible physical origins of the soft lags.

\subsection{Reverberation lag}
It is well established that the soft lags in AGNs are ``reverberation lags'', i.e. the time lags between the primary hard X-ray radiation
emitted by hot coronae, and the reflection radiation of the accretion disc due to primary radiation irradiating the disc.
\citet{de_marco_discovery_2013} performed a systematic study of the reverberation lags in AGNs, and found
correlations between the amplitude of the reverberation lag and the mass of the super-massive black hole,
and between the variability frequency and the black hole mass.
The authors found that, between 0.3$-$1 keV and 1$-$5 keV,
\begin{eqnarray}
{\rm log_{10}} \ \nu_{\rm lag} &=& -3.50\pm0.07 - 0.47\pm0.09\ {\rm log_{10}}\ M_7,\\ 
 {\rm log_{10}}\ |\tau| &=& 1.98\pm0.08 + 0.59\pm 0.11\ {\rm log_{10}}\ M_7,
\end{eqnarray}
where $\nu_{\rm lag}$ is the frequency of the reverberation lag in Hz, 
$\tau$ is the amplitude of the reverberation lag in second, 
and $M_7$ is the mass of the super-massive black hole in $10^7~\rm M_\odot$.
Substituting the black hole mass of \src{} \citep[${\rm log_{10}}~ M_{\rm BH}=7.7\pm0.4$;][]{wevers_fainter_2020},
we obtain ${\rm log_{10}}\ \nu_{\rm lag}=-3.83\pm0.21$, and ${\rm log_{10}} \ |\tau|=2.39\pm0.26$.
For comparison, the soft lags in \src{} are detected at $\frezero{}$, and the amplitude of the time lags between 0.3$-$1
and 1$-$5 keV is expected to be $-3.74\pm0.53~\rm ks$ (using Eq.~\ref{eq:logfit}).
Either the variability frequency or the amplitude is inconsistent with the reverberation lag in AGNs. This indicates that the soft lags in 
\src{} are not reverberation lags. The reverberation lag scenario is also disfavored by the absence of soft excess in the
covariance spectrum \citep{uttley_causal_2011}.

\subsection{Scattering}
Alternatively, the soft lags could be due to hard photons getting down-scattered by the materials in the outflow.
The unification models \citep[e.g.][]{dai_unified_2018} predicted that, at intermediate inclinations, 
the observers are still expected to see X-ray emission that originates from the outflow down-scattering hard X-ray photons.
Naturally we expect to see soft lags as on average the photons lose energy by down-scattering.

In the Compton down-scattering process, the average energy loss per scattering is
\begin{equation}
\frac{<x^\prime - x>}{x} = -x,\quad {\rm if}~kT_e/m_e c^2 \ll x \ll 1,
\end{equation}
where $x^\prime\equiv E^\prime/m_e c^2$ and $x\equiv E/m_e c^2$ are the dimensionless energies of the photons 
after and before the scattering, respectively, $m_e$ is the electron rest mass, and $kT_e$ is the temperature of the down-scattering medium.
For a photon with an energy of $x_0 (\ll 1)$, after $N$ scattering, the expected energy of the photon is
\begin{equation}
x_N \approx x_0 - N x_0^2.
\end{equation}
Therefore 
\begin{equation}
\frac{x_N}{x_0} \approx 1-Nx_0^2  \approx e^{-Nx_0^2},
\end{equation}
leading to 
\begin{equation}
 N \approx {\rm ln}\left(\frac{x_0}{x_N}\right) / x_0^2.
\end{equation}
Naturally we expect the amplitude of the time lags to be proportional to the number of scatterings.
Therefore in this down-scattering scenario the time lags between two energy bands are expected to be
proportional to the logarithm of the energy separation between two bands.
Thus this scenario is also consistent with the observations that the amplitude of the lag depends logarithmically on the energy separation
between two bands (Fig.~\ref{fig:lagvse}). 

It is interesting to note that soft lags with amplitudes of a few hundreds to thousands
of seconds have been detected in four ULXs: NGC 55 ULX1 \citep{pinto_ultraluminous_2017}, 
NGC 1313 X1 \citep{kara_discovery_2020}, NGC 4559 X7 \citep{pintore_rare_2021}, and NGC 7456 ULX-1 \citep{mondal_evidence_2021}.
For the lag amplitude to be
consistent with the reverberation lags in AGNs, the black hole masses have to be at least a few times $10^7~\rm M_\odot$
\citep{de_marco_discovery_2013}, much too massive for ULXs: actually
most ULXs are believed to contain stellar-mass black holes or neutron stars instead of IMBHs.
These time lags have been interpreted as due to hard photons getting down-scattered by outflows, 
the same scenario with what we use to explain the soft lags in \src. As both TDEs and ULXs are accreting
at super-Eddington rates where launching of outflow is expected, it is not surprising that
we detect soft lags due to the same physical process in these two kinds of objects.

\section*{acknowledgements}
The author is grateful to the anonymous referee for his/her very useful suggestions.
The author would like to thank Iossif Papadakis and Chichuan Jin for helpful discussion.
The author acknowledges the support by the Strategic Pioneer Program on Space Science, Chinese Academy of Sciences through grant XDA15052100.
This research has made use of data obtained through the High Energy Astrophysics Science Archive Research Center Online Service,
provided by the NASA/Goddard Space Flight Center.
This research makes uses of \textsc{Matplotlib} \citep{hunter_matplotlib:_2007},
a Python 2D plotting library which produces publication quality figures.

\section*{Data Availability}
The data underlying this article are publicly available from NASA's High Energy Astrophysics Science Archive
Research Center (HEASARC) archive at https://heasarc.gsfc.nasa.gov and
ESA's XMM-Newton Science Archive (XSA) at https://www.cosmos.esa.int/web/xmm-newton/xsa.

\bibliographystyle{mnras}
\bibliography{at2018fyk}

\begin{thebibliography}{}
\makeatletter
\relax
\def\mn@urlcharsother{\let\do\@makeother \do\$\do\&\do\#\do\^\do\_\do\%\do\~}
\def\mn@doi{\begingroup\mn@urlcharsother \@ifnextchar [ {\mn@doi@}
  {\mn@doi@[]}}
\def\mn@doi@[#1]#2{\def\@tempa{#1}\ifx\@tempa\@empty \href
  {http://dx.doi.org/#2} {doi:#2}\else \href {http://dx.doi.org/#2} {#1}\fi
  \endgroup}
\def\mn@eprint#1#2{\mn@eprint@#1:#2::\@nil}
\def\mn@eprint@arXiv#1{\href {http://arxiv.org/abs/#1} {{\tt arXiv:#1}}}
\def\mn@eprint@dblp#1{\href {http://dblp.uni-trier.de/rec/bibtex/#1.xml}
  {dblp:#1}}
\def\mn@eprint@#1:#2:#3:#4\@nil{\def\@tempa {#1}\def\@tempb {#2}\def\@tempc
  {#3}\ifx \@tempc \@empty \let \@tempc \@tempb \let \@tempb \@tempa \fi \ifx
  \@tempb \@empty \def\@tempb {arXiv}\fi \@ifundefined
  {mn@eprint@\@tempb}{\@tempb:\@tempc}{\expandafter \expandafter \csname
  mn@eprint@\@tempb\endcsname \expandafter{\@tempc}}}

\bibitem[\protect\citeauthoryear{Alston, Done  \& Vaughan}{Alston
  et~al.}{2014}]{alston_X-ray_2014}
Alston W.~N.,  Done C.,   Vaughan S.,  2014, \mn@doi [MNRAS]
  {10.1093/mnras/stu005}, 439, 1548

\bibitem[\protect\citeauthoryear{Altamirano \& M{\'e}ndez}{Altamirano \&
  M{\'e}ndez}{2015}]{altamirano_evolution_2015}
Altamirano D.,  M{\'e}ndez M.,  2015, \mn@doi [MNRAS] {10.1093/mnras/stv556},
  449, 4027

\bibitem[\protect\citeauthoryear{Ar{\'e}valo, Papadakis, Uttley, McHardy  \&
  Brinkmann}{Ar{\'e}valo et~al.}{2006}]{arevalo_spectral-timing_2006}
Ar{\'e}valo P.,  Papadakis I.~E.,  Uttley P.,  McHardy I.~M.,   Brinkmann W.,
  2006, \mn@doi [MNRAS] {10.1111/j.1365-2966.2006.10871.x}, 372, 401

\bibitem[\protect\citeauthoryear{Ar{\'e}valo, McHardy  \& Summons}{Ar{\'e}valo
  et~al.}{2008}]{arevalo_X-ray_2008}
Ar{\'e}valo P.,  McHardy I.~M.,   Summons D.~P.,  2008, \mn@doi [MNRAS]
  {10.1111/j.1365-2966.2008.13367.x}, 388, 211

\bibitem[\protect\citeauthoryear{Cackett, Fabian, Zogbhi, Kara, Reynolds  \&
  Uttley}{Cackett et~al.}{2013}]{cackett_soft_2013}
Cackett E.~M.,  Fabian A.~C.,  Zogbhi A.,  Kara E.,  Reynolds C.,   Uttley P.,
  2013, \mn@doi [ApJ] {10.1088/2041-8205/764/1/L9}, 764, L9

\bibitem[\protect\citeauthoryear{Crary, Finger, Kouveliotou, {van der Hooft},
  {van der Klis}, Lewin  \& {van Paradijs}}{Crary
  et~al.}{1998}]{crary_hard_1998}
Crary D.~J.,  Finger M.~H.,  Kouveliotou C.,  {van der Hooft} F.,  {van der
  Klis} M.,  Lewin W. H.~G.,   {van Paradijs} J.,  1998, \mn@doi [ApJL]
  {10.1086/311129}, 493, L71

\bibitem[\protect\citeauthoryear{Cui, Zhang  \& Chen}{Cui
  et~al.}{2000}]{cui_phase_2000}
Cui W.,  Zhang S.~N.,   Chen W.,  2000, \mn@doi [ApJL] {10.1086/312520}, 531,
  L45

\bibitem[\protect\citeauthoryear{Dai, McKinney, Roth, {Ramirez-Ruiz}  \&
  Miller}{Dai et~al.}{2018}]{dai_unified_2018}
Dai L.,  McKinney J.~C.,  Roth N.,  {Ramirez-Ruiz} E.,   Miller M.~C.,  2018,
  \mn@doi [ApJL] {10.3847/2041-8213/aab429}, 859, L20

\bibitem[\protect\citeauthoryear{De~Marco, Ponti, Cappi, Dadina, Uttley,
  Cackett, Fabian  \& Miniutti}{De~Marco
  et~al.}{2013a}]{de_marco_discovery_2013}
De~Marco B.,  Ponti G.,  Cappi M.,  Dadina M.,  Uttley P.,  Cackett E.~M.,
  Fabian A.~C.,   Miniutti G.,  2013a, \mn@doi [MNRAS] {10.1093/mnras/stt339},
  431, 2441

\bibitem[\protect\citeauthoryear{De~Marco, Ponti, Miniutti, Belloni, Cappi,
  Dadina  \& {Mu{\~n}oz-Darias}}{De~Marco et~al.}{2013b}]{de_marco_time_2013}
De~Marco B.,  Ponti G.,  Miniutti G.,  Belloni T.,  Cappi M.,  Dadina M.,
  {Mu{\~n}oz-Darias} T.,  2013b, \mn@doi [MNRAS] {10.1093/mnras/stt1853}, 436,
  3782

\bibitem[\protect\citeauthoryear{De~Marco, Ponti, {Mu{\~n}oz-Darias}  \&
  Nandra}{De~Marco et~al.}{2015}]{de_marco_tracing_2015}
De~Marco B.,  Ponti G.,  {Mu{\~n}oz-Darias} T.,   Nandra K.,  2015, \mn@doi
  [ApJ] {10.1088/0004-637X/814/1/50}, 814, 50

\bibitem[\protect\citeauthoryear{De~Marco et~al.,}{De~Marco
  et~al.}{2017}]{de_marco_evolution_2017}
De~Marco B.,  et~al., 2017, \mn@doi [MNRAS] {10.1093/mnras/stx1649}, 471, 1475

\bibitem[\protect\citeauthoryear{Epitropakis \& Papadakis}{Epitropakis \&
  Papadakis}{2016}]{epitropakis_statistical_2016}
Epitropakis A.,  Papadakis I.~E.,  2016, \mn@doi [A\&A]
  {10.1051/0004-6361/201527665}, 591, A113

\bibitem[\protect\citeauthoryear{Epitropakis \& Papadakis}{Epitropakis \&
  Papadakis}{2017}]{epitropakis_X-ray_2017}
Epitropakis A.,  Papadakis I.~E.,  2017, \mn@doi [MNRAS]
  {10.1093/mnras/stx612}, 468, 3568

\bibitem[\protect\citeauthoryear{Fabian et~al.,}{Fabian
  et~al.}{2009}]{fabian_broad_2009}
Fabian A.~C.,  et~al., 2009, \mn@doi [Nature] {10.1038/nature08007}, 459, 540

\bibitem[\protect\citeauthoryear{Fabian et~al.,}{Fabian
  et~al.}{2013}]{fabian_long_2013}
Fabian A.~C.,  et~al., 2013, \mn@doi [MNRAS] {10.1093/mnras/sts504}, 429, 2917

\bibitem[\protect\citeauthoryear{Gezari}{Gezari}{2021}]{gezari_tidal_2021}
Gezari S.,  2021, arXiv e-prints, p. arXiv:2104.14580

\bibitem[\protect\citeauthoryear{Guillochon \& {Ramirez-Ruiz}}{Guillochon \&
  {Ramirez-Ruiz}}{2015}]{guillochon_dark_2015}
Guillochon J.,  {Ramirez-Ruiz} E.,  2015, \mn@doi [ApJ]
  {10.1088/0004-637X/809/2/166}, 809, 166

\bibitem[\protect\citeauthoryear{Heil \& Vaughan}{Heil \&
  Vaughan}{2010}]{heil_linear_2010}
Heil L.~M.,  Vaughan S.,  2010, \mn@doi [MNRAS]
  {10.1111/j.1745-3933.2010.00864.x}, 405, L86

\bibitem[\protect\citeauthoryear{{Hern{\'a}ndez-Garc{\'i}a}, Vaughan, Roberts
  \& Middleton}{{Hern{\'a}ndez-Garc{\'i}a}
  et~al.}{2015}]{hernandez-garcia_X-ray_2015}
{Hern{\'a}ndez-Garc{\'i}a} L.,  Vaughan S.,  Roberts T.~P.,   Middleton M.,
  2015, \mn@doi [MNRAS] {10.1093/mnras/stv1830}, 453, 2877

\bibitem[\protect\citeauthoryear{Hunter}{Hunter}{2007}]{hunter_matplotlib:_2007}
Hunter J.~D.,  2007, \mn@doi [Computing in Science Engineering]
  {10.1109/MCSE.2007.55}, 9, 90

\bibitem[\protect\citeauthoryear{Kara et~al.,}{Kara
  et~al.}{2020}]{kara_discovery_2020}
Kara E.,  et~al., 2020, \mn@doi [MNRAS] {10.1093/mnras/stz3318}, 491, 5172

\bibitem[\protect\citeauthoryear{McHardy, Papadakis, Uttley, Page  \&
  Mason}{McHardy et~al.}{2004}]{mchardy_combined_2004}
McHardy I.~M.,  Papadakis I.~E.,  Uttley P.,  Page M.~J.,   Mason K.~O.,  2004,
  \mn@doi [MNRAS] {10.1111/j.1365-2966.2004.07376.x}, 348, 783

\bibitem[\protect\citeauthoryear{Miyamoto \& Kitamoto}{Miyamoto \&
  Kitamoto}{1989}]{miyamoto_X-ray_1989}
Miyamoto S.,  Kitamoto S.,  1989, \mn@doi [Nature] {10.1038/342773a0}, 342, 773

\bibitem[\protect\citeauthoryear{Miyamoto, Kitamoto, Mitsuda  \&
  Dotani}{Miyamoto et~al.}{1988}]{miyamoto_delayed_1988}
Miyamoto S.,  Kitamoto S.,  Mitsuda K.,   Dotani T.,  1988, \mn@doi [Nature]
  {10.1038/336450a0}, 336, 450

\bibitem[\protect\citeauthoryear{Mondal, R{\'o}{\.z}a{\'n}ska, De~Marco  \&
  Markowitz}{Mondal et~al.}{2021}]{mondal_evidence_2021}
Mondal S.,  R{\'o}{\.z}a{\'n}ska A.,  De~Marco B.,   Markowitz A.,  2021,
  \mn@doi [MNRAS] {10.1093/mnrasl/slab061}, 505, L106

\bibitem[\protect\citeauthoryear{Nowak \& Vaughan}{Nowak \&
  Vaughan}{1996}]{nowak_phase_1996}
Nowak M.~A.,  Vaughan B.~A.,  1996, \mn@doi [MNRAS] {10.1093/mnras/280.1.227},
  280, 227

\bibitem[\protect\citeauthoryear{Nowak, Vaughan, Wilms, Dove  \&
  Begelman}{Nowak et~al.}{1999}]{nowak_rossi_1999}
Nowak M.~A.,  Vaughan B.~A.,  Wilms J.,  Dove J.~B.,   Begelman M.~C.,  1999,
  \mn@doi [ApJ] {10.1086/306610}, 510, 874

\bibitem[\protect\citeauthoryear{Papadakis, Nandra  \& Kazanas}{Papadakis
  et~al.}{2001}]{papadakis_frequency-dependent_2001}
Papadakis I.~E.,  Nandra K.,   Kazanas D.,  2001, \mn@doi [ApJL]
  {10.1086/321722}, 554, L133

\bibitem[\protect\citeauthoryear{Papadakis, Rigas, Markowitz  \&
  McHardy}{Papadakis et~al.}{2019}]{papadakis_long_2019}
Papadakis I.~E.,  Rigas A.,  Markowitz A.,   McHardy I.~M.,  2019, \mn@doi
  [MNRAS] {10.1093/mnras/stz489}, 485, 1454

\bibitem[\protect\citeauthoryear{Pinto et~al.,}{Pinto
  et~al.}{2017}]{pinto_ultraluminous_2017}
Pinto C.,  et~al., 2017, \mn@doi [MNRAS] {10.1093/mnras/stx641}, 468, 2865

\bibitem[\protect\citeauthoryear{Pintore et~al.,}{Pintore
  et~al.}{2021}]{pintore_rare_2021}
Pintore F.,  et~al., 2021, \mn@doi [MNRAS] {10.1093/mnras/stab913}, 504, 551

\bibitem[\protect\citeauthoryear{Reig, Kylafis  \& Giannios}{Reig
  et~al.}{2003}]{reig_energy_2003}
Reig P.,  Kylafis N.~D.,   Giannios D.,  2003, \mn@doi [A\&A]
  {10.1051/0004-6361:20030449}, 403, L15

\bibitem[\protect\citeauthoryear{Sriram, Agrawal  \& Rao}{Sriram
  et~al.}{2009}]{sriram_energy-dependent_2009}
Sriram K.,  Agrawal V.~K.,   Rao A.~R.,  2009, \mn@doi [ApJ]
  {10.1088/0004-637X/700/2/1042}, 700, 1042

\bibitem[\protect\citeauthoryear{Stanek}{Stanek}{2018}]{stanek_asas-sn_2018}
Stanek K.~Z.,  2018, Transient Name Server Discovery Report, 2018--1325, 1

\bibitem[\protect\citeauthoryear{Timmer \& Koenig}{Timmer \&
  Koenig}{1995}]{timmer_generating_1995}
Timmer J.,  Koenig M.,  1995, A\&A, 300, 707

\bibitem[\protect\citeauthoryear{Uttley, Wilkinson, Cassatella, Wilms,
  Pottschmidt, Hanke  \& B{\"o}ck}{Uttley et~al.}{2011}]{uttley_causal_2011}
Uttley P.,  Wilkinson T.,  Cassatella P.,  Wilms J.,  Pottschmidt K.,  Hanke
  M.,   B{\"o}ck M.,  2011, \mn@doi [MNRAS] {10.1111/j.1745-3933.2011.01056.x},
  414, L60

\bibitem[\protect\citeauthoryear{Wevers}{Wevers}{2020}]{wevers_fainter_2020}
Wevers T.,  2020, \mn@doi [MNRAS] {10.1093/mnrasl/slaa097}, 497, L1

\bibitem[\protect\citeauthoryear{Wevers et~al.,}{Wevers
  et~al.}{2019}]{wevers_evidence_2019}
Wevers T.,  et~al., 2019, \mn@doi [MNRAS] {10.1093/mnras/stz1976}, 488, 4816

\bibitem[\protect\citeauthoryear{Wevers et~al.,}{Wevers
  et~al.}{2021}]{wevers_rapid_2021}
Wevers T.,  et~al., 2021, \mn@doi [ApJ] {10.3847/1538-4357/abf5e2}, 912, 151

\bibitem[\protect\citeauthoryear{Wilkinson \& Uttley}{Wilkinson \&
  Uttley}{2009}]{wilkinson_accretion_2009}
Wilkinson T.,  Uttley P.,  2009, \mn@doi [MNRAS]
  {10.1111/j.1365-2966.2009.15008.x}, 397, 666

\bibitem[\protect\citeauthoryear{Zoghbi \& Fabian}{Zoghbi \&
  Fabian}{2011}]{zoghbi_X-ray_2011}
Zoghbi A.,  Fabian A.~C.,  2011, \mn@doi [MNRAS]
  {10.1111/j.1365-2966.2011.19655.x}, 418, 2642

\makeatother
\end{thebibliography}

\section*{Appendix}
To estimate the significance and uncertainty of the measured time-lags, for each energy band
(out of 0.3$-$0.5 keV, 0.5$-$1 keV, 1$-$2 keV, 2$-$3 keV, and 3$-$5 keV),
we fit the Leahy-normalized power spectral density with the following model:
\begin{equation}
 P(\nu) = N \nu^{-\alpha} + 2,
\end{equation}
i.e. a powerlaw function plus a constant power of 2.
The best-fit parameters and the covariance matrix are obtained by minimizing $\chi^2$. 
Then, we perform 10,000 simulations. In each simulation, for every energy band 
we sample a powerlaw power spectral density using the best-fit model parameters and the covariance matrix,
simulate the lightcurve following \citet{timmer_generating_1995}, and add Poisson noise.
The phase of the Fourier transform is kept to the the same with that of the reference band to ensure zero time-lag
between different energy bands. For the simulations we take a time resolution of 30 s, the same with the real lightcurves.
To avoid red noise leakage, the duration of the simulated lightcurves is $\sim180~\rm ks$, 5 times longer than the real lightcurves.
We randomly select a $\sim 30~\rm ks$ interval within the whole $\sim180~\rm ks$ interval,
and apply identical procedures as in Sec.~\ref{sec:data_reduction} to the $30~\rm ks$ time series to
compute the time-lags for the four energy bands. Finally we derive the confidence levels and uncertainties from the
distribution of the simulated time-lags.

\label{lastpage}
\end{document}